\documentclass[10pt,twocolumn]{article}

\usepackage[letterpaper,top=0.72in,bottom=0.78in,left=0.72in,right=0.72in,
            columnsep=0.28in]{geometry}
\usepackage[T1]{fontenc}
\usepackage{lmodern}
\usepackage{amsmath,amssymb,bm}
\usepackage{booktabs,array,multirow,threeparttable}
\usepackage{graphicx}
\usepackage[numbers,sort&compress]{natbib}
\usepackage{microtype}
\usepackage{titlesec}
\usepackage[font=small,labelfont=bf]{caption}
\usepackage{xcolor}
\usepackage{xurl}
\usepackage{hyperref}
\usepackage{placeins}
\usepackage{flushend}
\usepackage{bm}

\hypersetup{
  colorlinks=true,
  linkcolor=blue!45!black,
  citecolor=green!35!black,
  urlcolor=blue!55!black,
  pdftitle={Alpha-Core Breakup in the Strong Decay of Double-Lambda Helium-6 to a Deeply Bound H Dibaryon},
  pdfauthor={Mahboubeh Shahrbaf and Makoto Oka},
  pdfsubject={Form-factor revision}
}

\titleformat{\section}
  {\centering\normalfont\bfseries\small}
  {\Roman{section}.}{0.55em}{\MakeUppercase}
\titleformat{\subsection}
  {\centering\normalfont\bfseries\normalsize}
  {\Alph{subsection}.}{0.45em}{}
\titlespacing*{\section}{0pt}{1.6ex plus 0.5ex minus 0.2ex}{1.0ex}
\titlespacing*{\subsection}{0pt}{1.4ex plus 0.4ex minus 0.2ex}{0.8ex}

\newcommand{\HeLL}{\,{}_{\Lambda\Lambda}^{6}\mathrm{He}}
\newcommand{\HeFour}{\,{}^{4}\mathrm{He}}
\newcommand{\HeThree}{\,{}^{3}\mathrm{He}}
\newcommand{\tritium}{\,{}^{3}\mathrm{H}}
\newcommand{\Mcal}{\mathcal{M}_{0}}
\newcommand{\hbarc}{\hbar c}

\begin{document}

\makeatletter
\twocolumn[
\begin{@twocolumnfalse}
\begin{center}
{\LARGE\bfseries
Alpha-Core Breakup in the Strong Decay of
\(\HeLL\) to a Deeply Bound \(H\) Dibaryon\par}
\vspace{0.9em}
{\large
Mahboubeh Shahrbaf\({}^{1,2,*}\) and Makoto Oka\({}^{2,3,4}\)\par}
\vspace{0.55em}
{\small
\({}^{1}\)Institute of Theoretical Physics, University of Wroc{\l}aw,
50-204 Wroc{\l}aw, Poland\par
\({}^{2}\)Nishina Center for Accelerator-Based Science, RIKEN,
Wako 351-0198, Japan\par
\({}^{3}\)Department of Physics, Tohoku University,
Sendai 980-8578, Japan\par
\({}^{4}\)Advanced Science Research Center, Japan Atomic Energy Agency,
Tokai, Ibaraki 319-1195, Japan\par}
\vspace{0.45em}
{\footnotesize
\({}^{*}\)Corresponding author:
\href{mailto:m.shahrbaf46@gmail.com}{m.shahrbaf46@gmail.com}\par}
\vspace{0.85em}
\begin{minipage}{0.91\textwidth}
\small

\begin{abstract}
We investigate the effect of $\alpha$-core breakup on the strong conversion
of \(\HeLL\) into a deeply bound \(H\) dibaryon.  In addition to the coherent
\(H+\HeFour\) channel, we evaluate the open final states
\(H+p+\tritium\), \(H+n+\HeThree\), and \(H+d+d\) using a translationally
invariant Gaussian cluster description, including spin-isospin recoupling
and full nonrelativistic three-body phase-space integrations.  The breakup
widths are normalized to Gal's intact-\(\alpha\) result.  At
\(B_{\Lambda\Lambda}^{H}=176~\mathrm{MeV}\), corresponding to
\(m_H\simeq2055~\mathrm{MeV}\), the summed breakup width exceeds the
intact-\(\alpha\) width by a factor \(R_{\mathrm{br}}=2.49\times10^{3}\).
The resulting inclusive width and lifetime are
\(\Gamma_{\mathrm{inc}}=3.87\times10^{-4}~\mathrm{eV}\) and
\(\tau_{\mathrm{inc}}=1.70\times10^{-12}~\mathrm{s}\), respectively,
compared with \(\tau_{\alpha}=4.25\times10^{-9}~\mathrm{s}\) for the
intact-\(\alpha\) channel alone.  The mass-dependent calculation shows that
the inclusive lifetime crosses the characteristic hypernuclear weak-decay
timescale near \(m_H\simeq2020~\mathrm{MeV}\) and increases rapidly as the
\(H\) mass decreases.  In the representative dark-matter-motivated interval
\(1865\leq m_H\leq1885~\mathrm{MeV}\), we obtain
\(6.59\times10^{-4}\lesssim\tau_{\mathrm{inc}}\lesssim
6.80\times10^{-3}~\mathrm{s}\), far exceeding the weak-decay timescale.
Thus, although core breakup can dominate the inclusive strong width near
\(m_H\simeq2055~\mathrm{MeV}\), weakly decaying double-\(\Lambda\)
hypernuclei remain compatible, within the present framework, with a deeply
bound \(uuddss\) state in the mass range relevant to sexaquark dark matter.
\end{abstract}
\end{minipage}
\end{center}
\vspace{0.9em}
\end{@twocolumnfalse}
]
\makeatother

\section{Introduction}
\label{sec:introduction}

The $H$ dibaryon is one of the longest-standing candidates for an exotic multiquark state in quantum chromodynamics (QCD). It was originally proposed by Jaffe as a flavor-singlet six-quark configuration with quark content $uuddss$, baryon number $B=2$, strangeness $S=-2$, isospin $I=0$, and spin-parity $J^{P}=0^{+}$, with a mass approximately $80~\mathrm{MeV}$ below the $\Lambda\Lambda$ threshold in the MIT bag model \cite{Jaffe1977}. Early quark-cluster studies investigated the \(H\) as a compact dihyperon
state \cite{Oka:1983ku}, while the role of instanton-induced interactions
in determining its stability was examined by Takeuchi and Oka
\cite{Takeuchi:1990qj}; a comprehensive review of the early
theoretical developments
is given in Ref.~\cite{Sakai:1999qm}.
 Despite extensive theoretical and experimental study, the spectrum in this channel remains unsettled. Lattice-QCD calculations performed at heavier-than-physical quark masses or in the $SU(3)_{f}$-symmetric limit have found substantial attraction and, in several cases, a bound state \cite{Beane2011,Inoue2011,Green2021}. Chiral extrapolations and coupled-channel calculations closer to the physical point, however, generally indicate much weaker binding or an unbound near-threshold state \cite{Shanahan2011,Sasaki2020}. 
%A recent constituent-quark calculation further illustrates the strong dependence of the predicted mass and structure on how the six-quark Hilbert space and antisymmetrization are implemented \cite{Gordillo2026}. 
Experimental searches using the
\(^{12}\mathrm{C}(K^-,K^+)\) reaction include the KEK search for an
\(H\) dibaryon resonance \cite{Yoon:2007aq} and the most recent dedicated
search at J-PARC, which targets the regions near the \(\Lambda\Lambda\)
and \(\Xi^-p\) thresholds but whose final results have not yet been
reported \cite{Kim:2022bwb}. Complementary searches in bottomonium decays
and heavy-ion collisions have found no unambiguous signal in the mass and
lifetime regions to which they are sensitive
\cite{Kim2013,Adam2016,Lees2019}. Thus, although attraction in the $S=-2$, $I=0$ channel is well established theoretically, neither a near-threshold $H$ nor a deeply bound compact state has yet been demonstrated at the physical point.

Double-$\Lambda$ hypernuclei provide a qualitatively different and particularly important constraint. The NAGARA event established the particle stability and subsequent weak decay of ${}^{6}_{\Lambda\Lambda}\mathrm{He}$, with a measured double-$\Lambda$ binding energy $B_{\Lambda\Lambda}({}^{6}_{\Lambda\Lambda}\mathrm{He})=6.91\pm0.17~\mathrm{MeV}$ \cite{Ahn2013,Hiyama2018}. If the binding of an $H$ relative to the free $\Lambda\Lambda$ threshold exceeds this value, the strong conversion
\begin{equation}
{}^{6}_{\Lambda\Lambda}\mathrm{He}\longrightarrow H+{}^{4}\mathrm{He}
\label{eq:intact_alpha}
\end{equation}
is energetically allowed. The observation of the parent hypernucleus through its weak decay has therefore often been used to argue against an $H$ bound by more than approximately $7~\mathrm{MeV}$ \cite{Dalitz1989,Ahn2013}. This argument is compelling only if the strong conversion in Eq.~\eqref{eq:intact_alpha} is sufficiently rapid. Kinematic accessibility alone does not determine the rate: the transition amplitude also depends on the short-distance $\Lambda\Lambda\to H$ overlap, correlations in the initial hypernucleus, and the momentum transferred to the nuclear core. These dynamical effects become especially important when the final dibaryon is both deeply bound and spatially compact.

This observation connects the hypernuclear problem to the proposal of a compact sexaquark as dark matter. The sexaquark, conventionally denoted by $S$, has the same quark content
and global quantum numbers as the $H$, but it is hypothesized to be a
compact color-flavor-spin singlet with only a small overlap with the
physical $\Lambda\Lambda$, $N\Xi$, and $\Sigma\Sigma$ baryon-baryon
channels, rather than a spatially extended hadronic molecule dominated
by one or more of these channels \cite{Farrar2003,Farrar2022Overview,FarrarWintergerst2023}. These two structural pictures should not be regarded as interchangeable, even though the notation $H$ and $S$ is sometimes used for both in the literature \cite{FarrarWintergerst2023,Gordillo2026}. A dedicated laboratory search for a compact, long-lived sexaquark has
also been proposed using its production approximately at rest in
antiprotonic ${}^{3}\mathrm{He}$ atoms through the reaction
$\bar{p}+{}^{3}\mathrm{He}\rightarrow
S+K^{+}+K^{+}+\pi^{-}$, followed by missing-mass reconstruction
\cite{Doser:2023gls}. In a complementary high-energy production study,
the coalescence of a compact sexaquark from three diquarks was investigated
within the PACIAE parton--hadron cascade model for $pp$ collisions at
$\sqrt{s}=7~\mathrm{TeV}$, and its predicted yield was compared with that
of a molecular $H(\Lambda\Lambda)$ configuration
\cite{She:2025dqx}. In the present work, $H$ denotes the hypothetical $B=2$, $S=-2$ final dibaryon in the hypernuclear transition; the term sexaquark refers specifically to its proposed deeply bound compact realization.

Deep binding is a central, rather than incidental, condition of the dark-matter proposal because it controls which decay channels are kinematically open. For $m_{S}<m_{d}+m_{e}\simeq1876.1~\mathrm{MeV}$, the sexaquark is stable against baryon-number-conserving decays, whereas in the interval above this threshold but below the lowest singly weak threshold, $m_{S}<m_{\Lambda}+m_{p}+m_{e}\simeq2054.5~\mathrm{MeV}$, its decay requires a doubly strangeness-changing weak transition \cite{FarrarZaharijas2004,Farrar2022Overview,FarrarWang2023}. Conversely, the observed stability of nuclei disfavors sufficiently smaller masses and motivates a conventional lower boundary near $1.86~\mathrm{GeV}$ \cite{Gross2018,FarrarWang2023}. The frequently discussed dark-matter interval $m_{S}\simeq1.86$--$1.89~\mathrm{GeV}$ therefore corresponds to a binding of approximately $0.34$--$0.37~\mathrm{GeV}$ below the $\Lambda\Lambda$ threshold. A small overlap between such a compact state and two separated baryons has been invoked to suppress its formation, breakup, and weak decay \cite{FarrarZaharijas2004,FarrarWang2023}. Whether QCD produces both the required deep binding and the required suppression remains an open dynamical question.

Moreover, deep binding and longevity are necessary but not sufficient for a viable dark-matter candidate. The state must also be produced with the required relic abundance and remain consistent with terrestrial, cosmological, and astrophysical constraints. QCD-crossover production has been proposed as a mechanism linking the sexaquark abundance to the baryon abundance \cite{farrar2022,FarrarWangXu2020}, while thermal-reaction analyses find that efficient number-changing processes can instead drive the surviving abundance to a negligible value unless the relevant conversion amplitudes are extremely suppressed \cite{KolbTurner2019,Moore:2024mot}. A recent study has identified late nonthermal production as an alternative, explicitly model-dependent route \cite{Moore:2025zyg}. In dense matter, rapid production in proto-neutron star conditions and the softening associated with a bosonic dibaryon pose additional challenges \cite{McDermott2019}, while hybrid-star studies show that the macroscopic outcome is sensitive to in-medium interactions and to the possible onset of deconfined quark matter \cite{Shahrbaf2022, Shahrbaf:2023uxy, Shahrbaf:2024gdm}. More recently, color-spin molecular-dynamics calculations of
beta-equilibrated neutron star matter including strangeness have found
that color-magnetic interactions favor the self-consistent formation of
correlated color-singlet multiquark clusters, with cluster sizes
preferentially occurring at quark numbers that are multiples of three
\cite{Yasutake:2026fjh}. Although individual clusters could not be
identified unambiguously with particular exotic hadrons, that study
explicitly noted sexaquarks among the possible multiquark configurations
relevant to dense matter. Taken together, the results reported in the literature motivate testing each component of the scenario separately. Accordingly, the present work addresses the hypernuclear constraint on \(H\)-dibaryon formation.

Gal recently revisited the hypernuclear part of the problem in a realistic $\Lambda$--$\Lambda$--$\alpha$ three-body description of ${}^{6}_{\Lambda\Lambda}\mathrm{He}$ \cite{Gal2024}. For the coherent channel in Eq.~\eqref{eq:intact_alpha}, the large recoil required to produce a deeply bound $H$ leads to a strong spatial-overlap suppression. At the representative binding $B_{\Lambda\Lambda}^{H}=176~\mathrm{MeV}$, corresponding to $m_{H}\simeq2055~\mathrm{MeV}$ near the $\Lambda+n$ threshold, Gal obtained $\Gamma_{\alpha}=1.55\times10^{-7}~\mathrm{eV}$ and $\tau_{\alpha}=4.245\times10^{-9}~\mathrm{s}$. This is longer than the characteristic weak-decay timescale, $\tau_{\mathrm{weak}}\sim10^{-10}~\mathrm{s}$, of the parent double-$\Lambda$ hypernucleus. More generally, the intact-$\alpha$ lifetime was found to increase rapidly as $m_{H}$ decreases, showing that the observation of weakly decaying double-$\Lambda$ hypernuclei does not by itself exclude a sufficiently deeply bound $H$ \cite{Gal2024}. In a separate calculation, Gal estimated the doubly weak decay $H\to nn$ and obtained a lifetime of order $10^{5}~\mathrm{s}$ in the relevant mass range, far below a cosmological lifetime \cite{Donoghue1986,FarrarZaharijas2004,Gal2024}. That conclusion concerns the lifetime of the produced $H$ itself and is logically distinct from the strong-decay lifetime of the parent hypernucleus studied here.

The coherent $H+{}^{4}\mathrm{He}$ channel, however, need not exhaust the inclusive strong width. At Gal's $B_{\Lambda\Lambda}^{H}=176~\mathrm{MeV}$ benchmark, the energy release in Eq.~\eqref{eq:intact_alpha} is approximately $169~\mathrm{MeV}$, large enough to disrupt the compact $\alpha$ core. The lowest charge-conserving breakup channels are
\begin{align}
{}^{6}_{\Lambda\Lambda}\mathrm{He}&\longrightarrow H+p+{}^{3}\mathrm{H},
\label{eq:breakup_pt}\\
{}^{6}_{\Lambda\Lambda}\mathrm{He}&\longrightarrow H+n+{}^{3}\mathrm{He},
\label{eq:breakup_nhe3}\\
{}^{6}_{\Lambda\Lambda}\mathrm{He}&\longrightarrow H+d+d.
\label{eq:breakup_dd}
\end{align}
Their breakup thresholds are only approximately $20$--$24~\mathrm{MeV}$, leaving about $145$--$149~\mathrm{MeV}$ of kinetic energy at this benchmark. In contrast to the two-body final state, the available energy can be shared between two independent Jacobi motions. The momentum-space suppression is therefore governed by a two-variable form factor rather than by the single coherent recoil momentum of the intact-$\alpha$ channel. The additional kinematic freedom and three-body phase space can compensate, at least partly, for the cost of breaking the $\alpha$ core and may substantially alter the inclusive conversion rate.

In this work, we extend Gal's intact-$\alpha$ calculation by evaluating the
three open $\alpha$-breakup channels introduced above. Their
two-Jacobi-momentum form factors are calculated within a translationally
invariant Gaussian cluster model and integrated over the complete
nonrelativistic three-body phase space. By normalizing the breakup widths
to Gal's intact-$\alpha$ result, the common short-distance
$\Lambda\Lambda\to H$ matrix element cancels within the present framework.

At the $B_{\Lambda\Lambda}^{H}=176~\mathrm{MeV}$ benchmark, the calculated
total breakup width exceeds the intact-$\alpha$ width by a factor of
approximately $2.49\times10^{3}$, yielding an inclusive lifetime of
$\tau_{\mathrm{inc}}\simeq1.70\times10^{-12}~\mathrm{s}$. Extending the
calculation over the $H$-mass range covered by Gal's tabulated results, we
find that the inclusive lifetime crosses the characteristic weak-decay
timescale near $m_H\simeq2020~\mathrm{MeV}$ and increases rapidly toward
lower masses. Thus, although $\alpha$ breakup is important near the
$\Lambda+n$ benchmark, it does not exclude the dark-matter-motivated
low-mass interval through the observed weak decay of
${}^{6}_{\Lambda\Lambda}\mathrm{He}$. This conclusion concerns the survival
of the parent hypernucleus and does not determine the intrinsic lifetime
or cosmological viability of the $H$ dibaryon.

The remainder of the paper is organized as follows. Section~II presents the transition amplitude, the intrinsic Gaussian construction, the cluster overlaps, and the three-body phase-space formalism. Section~III gives the benchmark numerical results and examines their structural sensitivity. Section~IV compares the inclusive calculation with Gal's intact-$\alpha$ results over the scanned $H$-mass range. Section~V summarizes the conclusions and the limitations of the present model.

\section{Method and formalism}
\label{sec:method}

\subsection{Transition amplitude and general breakup form factor}
\label{subsec:general-form-factor}

Consider a breakup partition \(X=a+b\).  We denote by
\(\bm r_{\Lambda\Lambda}\) the relative coordinate of the two
\(\Lambda\) hyperons, by \(\bm R\) their center-of-mass coordinate relative
to the alpha core, by \(\bm r_{ab}\) the relative coordinate between the
centers of mass of the two nuclear fragments, and by \(\bm\xi_a\) and
\(\bm\xi_b\) complete sets of intrinsic Jacobi coordinates of the
fragments.  The integration symbol \(d\xi_c\) below denotes integration over
all \(A_c-1\) three-dimensional intrinsic Jacobi vectors of cluster \(c\),
or equivalently over \(3(A_c-1)\) scalar dimensions.

Separating the spatial and spin-isospin parts, the initial state is written
schematically as
\begin{equation}
 \Psi_i^{\mathrm{sp}}
 =
 \widetilde{\phi}_{\Lambda\Lambda}(\bm r_{\Lambda\Lambda})
 \Phi_{\Lambda\Lambda}(\bm R)
 \Psi_{\alpha}^{\mathrm{int}}
   (\bm\xi_a,\bm\xi_b,\bm r_{ab}),
 \label{eq:initial-factorization}
\end{equation}
together with the alpha-core spin-isospin state
\(\chi_{00}^{SI}\).  The tilde denotes the short-range correlation in the
initial \(\Lambda\Lambda\) component.  In the plane-wave approximation for
the two final Jacobi motions, the spatial final state is
\begin{equation}
 \begin{aligned}
 \Psi_{f,X}^{\mathrm{sp}}(\bm p,\bm q)
 ={}&
 \phi_H(\bm r_{\Lambda\Lambda})
 \Phi_a(\bm\xi_a)\Phi_b(\bm\xi_b)
 \\
 &\times
 e^{-i\bm p\cdot\bm r_{ab}}
 e^{-i\bm q\cdot\bm R},
 \end{aligned}
 \label{eq:general-final-state}
\end{equation}
where \(\phi_H\) is the compact-\(H\) wave function,
\(\bm p\) is the \(a\)-\(b\) relative Jacobi momentum, and \(\bm q\) is the
momentum of \(H\) relative to the center of mass of the \(ab\) pair.
Indeed, $\phi_H$, $\Phi_a$, and $\Phi_b$ describe the internal bound
states of the outgoing composite fragments, whereas the last part describes the relative continuum motion of their
centers of mass.  In the present plane-wave approximation, interactions
among the outgoing fragments are neglected, and hence the final-state relative motion is treated as free.  
%The functions $\Phi_a$ and $\Phi_b$ do not provide any additional confinement of the outgoing particles; they describe only the internal structure of the composite fragments. For a single-nucleon fragment, no internal coordinate is present and the corresponding internal factor is set to unity.

For a transition operator \(V_{\Lambda\Lambda\rightarrow H}\), the decay
amplitude to a physical charge $X$ and spin state \(\lambda\) is
\begin{equation}
 \mathcal M_{X\lambda}(\bm p,\bm q)
 =
 \left\langle
 \Psi_{f,X}^{\mathrm{sp}}\chi_{X\lambda}^{SI}
 \left|V_{\Lambda\Lambda\rightarrow H}\right|
 \Psi_i^{\mathrm{sp}}\chi_{00}^{SI}
 \right\rangle .
 \label{eq:general-transition-amplitude}
\end{equation}
We assume that the short-range conversion acts on
\(\bm r_{\Lambda\Lambda}\), while the alpha nucleons are spectators.  The
amplitude then separates as
\begin{equation}
 \mathcal M_{X\lambda}(\bm p,\bm q)
 =
 \Mcal\,c_{X\lambda}^{SI}\,F_X(\bm p,\bm q),
 \label{eq:amplitude-factorization}
\end{equation}
where
\begin{equation}
 \Mcal
 =
 \int d^3r_{\Lambda\Lambda}\,
 \phi_H^{*}(\bm r_{\Lambda\Lambda})
 V_{\Lambda\Lambda\rightarrow H}
 \widetilde{\phi}_{\Lambda\Lambda}(\bm r_{\Lambda\Lambda})
 \label{eq:short-range-amplitude}
\end{equation}
contains the common short-distance dynamics, and
\begin{equation}
 c_{X\lambda}^{SI}
 =
 \langle\chi_{X\lambda}^{SI}|\chi_{00}^{SI}\rangle
 \label{eq:si-amplitude-definition}
\end{equation}
is a spin-isospin Clebsch-Gordan amplitude.

Before any Gaussian or cluster-factorization assumption is made, the
spatial breakup form factor is
\begin{align}
 F_X(\bm p,\bm q)
 ={}&
 \int d\xi_a\,d\xi_b\,d^3r_{ab}\,d^3R\,
 \Phi_a^{*}(\bm\xi_a)\Phi_b^{*}(\bm\xi_b)
 \nonumber\\
 &\times
 e^{i\bm p\cdot\bm r_{ab}}
 e^{i\bm q\cdot\bm R}
 \Psi_{\alpha}^{\mathrm{int}}
   (\bm\xi_a,\bm\xi_b,\bm r_{ab})
 \Phi_{\Lambda\Lambda}(\bm R).
 \label{eq:general-breakup-form-factor}
\end{align}
Equation~\eqref{eq:general-breakup-form-factor} is the general overlap
corresponding to the two independent final Jacobi momenta.  In particular,
it does not require the \(\bm p\) and \(\bm q\) dependences to factorize.

\subsection{Factorized form factor and Gal normalization}
\label{subsec:factorization}

We assume that the component of the intrinsic alpha wave
function associated with partition \(X=a+b\) can be approximated by
\begin{equation}
 \Psi_{\alpha}^{\mathrm{int}}
   (\bm\xi_a,\bm\xi_b,\bm r_{ab})
 \simeq
 \Phi_a^{(\alpha)}(\bm\xi_a)
 \Phi_b^{(\alpha)}(\bm\xi_b)
 \Phi_{\mathrm{rel}}^{(X)}(\bm r_{ab}).
 \label{eq:alpha-cluster-factorization}
\end{equation}
Introduce the two internal cluster-overlap amplitudes
\begin{equation}
 \mathcal O_c
 \equiv
 \int d\xi_c\,
 \Phi_c^{*}(\bm\xi_c)\Phi_c^{(\alpha)}(\bm\xi_c).
 \label{eq:oc-general}
\end{equation}
Substitution into Eq.~\eqref{eq:general-breakup-form-factor} then gives
four independent overlap integrals:
\begin{align}
 F_X(\bm p,\bm q)
 ={}&
 \left[
 \int d\xi_a\,
 \Phi_a^{*}(\bm\xi_a)\Phi_a^{(\alpha)}(\bm\xi_a)
 \right]
 \nonumber\\
 &\times
 \left[
 \int d\xi_b\,
 \Phi_b^{*}(\bm\xi_b)\Phi_b^{(\alpha)}(\bm\xi_b)
 \right]
 \nonumber\\
 &\times
 \left[
 \int d^3r_{ab}\,
 e^{i\bm p\cdot\bm r_{ab}}
 \Phi_{\mathrm{rel}}^{(X)}(\bm r_{ab})
 \right]
 \nonumber\\
 &\times
 \left[
 \int d^3R\,
 e^{i\bm q\cdot\bm R}
 \Phi_{\Lambda\Lambda}(\bm R)
 \right].
 \label{eq:four-factor-overlap}
\end{align}
The first two factors in Eq.~\eqref{eq:four-factor-overlap} are the
internal overlap integrals defined generally in
Eq.~\eqref{eq:oc-general}.  Defining
\begin{align}
 \mathcal C_X
 &\equiv\mathcal O_a\mathcal O_b,
 \label{eq:cx-general}\\
 F_{\mathrm{rel}}^{(X)}(\bm p)
 &\equiv
 \int d^3r_{ab}\,
 e^{i\bm p\cdot\bm r_{ab}}
 \Phi_{\mathrm{rel}}^{(X)}(\bm r_{ab}),
 \label{eq:frel-general}\\
 F_R(\bm q)
 &\equiv
 \int d^3R\,
 e^{i\bm q\cdot\bm R}
 \Phi_{\Lambda\Lambda}(\bm R),
 \label{eq:fr-general}
\end{align}
the form factor becomes
\begin{equation}
 F_X(\bm p,\bm q)
 =
 \mathcal C_X
 F_{\mathrm{rel}}^{(X)}(\bm p)F_R(\bm q)
 \equiv
 F_{\mathrm{breakup}}^{(X)}(\bm p)F_R(\bm q).
 \label{eq:factorized-breakup-form-factor}
\end{equation}
Here
\(F_{\mathrm{breakup}}^{(X)}
=\mathcal C_XF_{\mathrm{rel}}^{(X)}\) contains both the internal
cluster overlaps and the alpha-core breakup motion.  Consequently,
\begin{equation}
 |F_X(\bm p,\bm q)|^2
 =
 |\mathcal C_X|^2
 |F_{\mathrm{rel}}^{(X)}(\bm p)|^2
 |F_R(\bm q)|^2 .
 \label{eq:squared-factorized-form-factor}
\end{equation}
This is the factorized form-factor assumption used throughout the
calculation.  %A microscopic breakup wave function could correlate the two Jacobi motions and would not, in general, reduce Eq.~\eqref{eq:general-breakup-form-factor} to Eq.~\eqref{eq:factorized-breakup-form-factor}.

The spin-summed squared amplitude follows directly:
\begin{align}
 \sum_{\lambda}|\mathcal M_{X\lambda}|^2
 ={}&
 |\Mcal|^2\mathcal P_X^{SI}
 |\mathcal C_X|^2
 |F_{\mathrm{rel}}^{(X)}(\bm p)|^2
 |F_R(\bm q)|^2,
 \label{eq:general-spin-summed-amplitude}\\
 \mathcal P_X^{SI}
 \equiv{}&
 \sum_{\lambda}|c_{X\lambda}^{SI}|^2.
 \label{eq:psi-definition}
\end{align}
Thus the spatial form factor, the spin-isospin projection, and the
short-distance transition amplitude are distinct ingredients.
%The three-body phase-space measure is a fourth, separate ingredient; it is introduced only in Eq.~\eqref{eq:dx-sixdim} and is not absorbed into \(F_X\), \(c_{X\lambda}^{SI}\), or \(\Mcal\).

For the intact-$\alpha$ channel, the $\alpha$ core is unchanged, so the only
long-distance spatial factor is \(F_R(\bm k_{\alpha})\).  With Gal's angular
integration convention,
\begin{equation}
 I(k_{\alpha},a_{\Phi})
 \equiv
 4\pi|F_R(\bm k_{\alpha})|^2.
 \label{eq:gal-I-definition}
\end{equation}
Gal's width is therefore written as \cite{Gal2024}
\begin{equation}
 \Gamma_{\alpha}=|\Mcal|^{2}D_{\alpha}^{\mathrm{Gal}},
 \qquad
 D_{\alpha}^{\mathrm{Gal}}
 =\frac{\mu_{H\alpha}k_{\alpha}}{(2\pi\hbarc)^2}
 I(k_{\alpha},a_{\Phi}),
 \label{eq:gal-D}
\end{equation}
where
\begin{equation}
 \mu_{H\alpha}=\frac{m_Hm_{\alpha}}{m_H+m_{\alpha}}
 \label{eq:muhalpha}
\end{equation}
is the \(H\)-alpha reduced mass and 
\begin{equation}
 I(k,a_{\Phi})
 =32\pi^{5/2}a_{\Phi}^{3}\exp(-a_{\Phi}^{2}k^{2}).
 \label{eq:gal-overlap}
\end{equation}
The parameter \(a_{\Phi}\) is the coordinate-space width of
\(\Phi_{\Lambda\Lambda}(\bm R)\).

The breakup widths are expressed in the same normalization,
\begin{equation}
 \Gamma_X=|\Mcal|^{2}D_X,
 \qquad X\in\{p+t,n+\HeThree,d+d\},
 \label{eq:breakup-factorization}
\end{equation}
so that the common short-distance strength cancels:
\begin{equation}
 R_X\equiv\frac{\Gamma_X}{\Gamma_{\alpha}}
 =\frac{D_X}{D_{\alpha}^{\mathrm{Gal}}},
 \qquad
 \Gamma_X=R_X\Gamma_{\alpha}.
 \label{eq:ratio-normalization}
\end{equation}
The common \(\Mcal\) in the intact and breakup channels is a central
dynamical assumption of the estimate.

\subsection{Intrinsic alpha Gaussian and center-of-mass removal}
\label{subsec:com-removal}

Having separated the spin-isospin state in
Eq.~\eqref{eq:general-transition-amplitude}, we begin with the spatial part
of four identical \(0s\) Gaussian orbitals in laboratory coordinates,
\begin{equation}
 \Psi_{\alpha}^{\mathrm{lab}}
 \propto
 \exp\!\left[-\frac{1}{2b_N^2}
 \sum_{i=1}^{4}\bm r_i^{\,2}\right],
 \label{eq:alpha-lab}
\end{equation}
where \(b_N\) is the nucleon-level Gaussian width.  The corresponding
normalized spin-isospin state is \(\chi_{00}^{SI}\), with total \(S=0\)
and \(T=0\).  Defining the alpha center of mass by
\begin{equation}
 \bm R_{\alpha}=\frac{1}{4}\sum_{i=1}^{4}\bm r_i,
 \label{eq:alpha-com}
\end{equation}
one has the exact identity
\begin{equation}
 \begin{aligned}
 \sum_{i=1}^{4}\bm r_i^{\,2}
 &=
 \sum_{i=1}^{4}(\bm r_i-\bm R_{\alpha})^{2}
 \\
 &\quad{}+4\bm R_{\alpha}^{\,2}.
 \end{aligned}
 \label{eq:com-identity}
\end{equation}
Consequently, Eq.~\eqref{eq:alpha-lab} factorizes into an intrinsic
wave function and a center-of-mass Gaussian:
\begin{align}
 \Psi_{\alpha}^{\mathrm{lab}}
 &=
 \Psi_{\alpha}^{\mathrm{int}}\Phi_{\mathrm{cm}}, \label{eq:lab-factor}\\
 \Psi_{\alpha}^{\mathrm{int}}
 &=
 \mathcal N_{\alpha}
 \exp\!\left[-\frac{1}{2b_N^2}
 \sum_{i=1}^{4}(\bm r_i-\bm R_{\alpha})^{2}\right],
 \label{eq:alpha-intrinsic}\\
 \Phi_{\mathrm{cm}}
 &\propto
 \exp\!\left[-\frac{\bm R_{\alpha}^{\,2}}{2B_{\mathrm{cm}}^{2}}\right],
 \qquad B_{\mathrm{cm}}=\frac{b_N}{2}.
 \label{eq:cm-gaussian}
\end{align}
Only \(\Psi_{\alpha}^{\mathrm{int}}\) is used below.

For an \(A\)-nucleon \(0s\) Gaussian with its center of mass removed, the
intrinsic rms radius satisfies
\begin{equation}
 r_A^{2}
 \equiv
 \frac{1}{A}\sum_{i=1}^{A}
 \left\langle(\bm r_i-\bm R_A)^2\right\rangle
 =
 \frac{A-1}{A}\frac{3}{2}b_A^{2}.
 \label{eq:rms-width}
\end{equation}
Using the experimental charge radius as a proxy for \(r_A\) gives
\begin{equation}
 b_A=
 \left[\frac{2A}{3(A-1)}\right]^{1/2}r_{\mathrm{ch}}(A),
 \label{eq:b-from-radius}
\end{equation}
where $b_A$ is the %width of the intrinsic $0s$ Gaussian wave function. 
coordinate-space
Gaussian width inferred from  charge radius applied to composite clusters with $A_c>1$.
For the alpha particle, the same prescription defines the
nucleon-level Gaussian width $b_N\equiv b_{\alpha}$,
\begin{equation}
 b_N=b_{\alpha}
 =\sqrt{\frac{8}{9}}\,r_{\mathrm{ch}}(\HeFour).
 \label{eq:bn-alpha}
\end{equation}

\subsection{\texorpdfstring{\(1+3\) and \(2+2\)}{1+3 and 2+2}
cluster decompositions}
\label{subsec:cluster-decomposition}

Consider a binary partition \(A=A_a+A_b\), with cluster center-of-mass
coordinates
\begin{equation}
 \bm R_a=\frac{1}{A_a}\sum_{i\in a}\bm r_i,
 \qquad
 \bm R_b=\frac{1}{A_b}\sum_{j\in b}\bm r_j,
 \label{eq:cluster-coms}
\end{equation}
and relative coordinate
\begin{equation}
 \bm r_{ab}=\bm R_a-\bm R_b.
 \label{eq:rab}
\end{equation}
The intrinsic coordinate sum separates exactly:
\begin{align}
 \sum_{i=1}^{A}(\bm r_i-\bm R_A)^2
 ={}&
 \sum_{i\in a}(\bm r_i-\bm R_a)^2
 +\sum_{j\in b}(\bm r_j-\bm R_b)^2
 \nonumber\\
 &+\frac{A_aA_b}{A}\bm r_{ab}^{\,2}.
 \label{eq:cluster-identity}
\end{align}
The last term in Eq.~\eqref{eq:cluster-identity} produces the cluster decomposition of the initial
alpha-particle wave function in the form of a Gaussian, which reads 
\begin{equation}
 \Phi_{\mathrm{rel}}^{(X)}(\bm r_{ab})
 =
 \frac{1}{(\pi b_{\mathrm{rel},X}^{2})^{3/4}}
 \exp\!\left[-\frac{\bm r_{ab}^{\,2}}
 {2b_{\mathrm{rel},X}^{2}}\right],
 \label{eq:relative-gaussian}
\end{equation}
with
\begin{equation}
 b_{\mathrm{rel},X}
 =b_N\sqrt{\frac{A}{A_aA_b}}.
 \label{eq:brel-general}
\end{equation}
Thus \(A_a+A_b=1+3\) applies to the \(p+t\) and
\(n+\HeThree\) channels, whereas \(A_a+A_b=2+2\) applies to \(d+d\).
The width in Eq.~\eqref{eq:brel-general} describes the cluster
distribution inside the initial alpha particle; it is not a confinement
length for the outgoing continuum fragments.

\subsection{Intrinsic overlaps of the final clusters}
\label{subsec:internal-overlap}

We now evaluate the internal overlaps \(\mathcal O_c\) defined generally
in Eq.~\eqref{eq:oc-general}.  The overlap of two normalized
one-dimensional Gaussians of widths \(b\) and \(c\) is
\begin{equation}
 \begin{aligned}
 \mathcal O_1(b,c)
 &=
 \int_{-\infty}^{\infty}dx\,
 \frac{e^{-x^2/(2b^2)}}{(\pi b^2)^{1/4}}
 \frac{e^{-x^2/(2c^2)}}{(\pi c^2)^{1/4}}
 \\
 &=
 \left(\frac{2bc}{b^2+c^2}\right)^{1/2}.
 \end{aligned}
 \label{eq:one-d-overlap}
\end{equation}
A cluster containing \(A_c\) nucleons has \(3(A_c-1)\) intrinsic
dimensions.  Its Gaussian overlap with the corresponding component of the
alpha wave function is therefore
\begin{equation}
 \mathcal O_c
 =
 \left(
 \frac{2b_cb_N}{b_c^2+b_N^2}
 \right)^{3(A_c-1)/2},
 \label{eq:cluster-overlap}
\end{equation}
where $b_c$ is the intrinsic Gaussian width defined in
Eq.~\eqref{eq:b-from-radius}. For a single proton or neutron,
$A_c=1$, there is no internal relative coordinate; therefore, no
cluster width $b_c$ is required and $\mathcal O_c=1$.
The general channel coefficient in Eq.~\eqref{eq:cx-general} and its
squared contribution are therefore
\begin{equation}
 \mathcal C_X=\mathcal O_a\mathcal O_b,
 \qquad
 |\mathcal C_X|^2=\mathcal O_a^2\mathcal O_b^2.
 \label{eq:channel-overlap}
\end{equation}

The recoil coordinate is described by the normalized Gaussian
\begin{equation}
 \Phi_{\Lambda\Lambda}(\bm R)
 =
 \frac{1}{(\pi a_{\Phi}^{2})^{3/4}}
 \exp\!\left(-\frac{R^2}{2a_{\Phi}^{2}}\right).
 \label{eq:recoil-coordinate-gaussian}
\end{equation}
For the Fourier convention
\begin{equation}
 \widetilde\phi_b(\bm k)
 =
 \int d^3r\,e^{i\bm k\cdot\bm r}
 \frac{e^{-r^2/(2b^2)}}{(\pi b^2)^{3/4}},
 \label{eq:fourier-convention}
\end{equation}
the normalized momentum-space factor is
\begin{equation}
 |\widetilde\phi_b(\bm k)|^2
 =
 8\pi^{3/2}b^3e^{-b^2k^2}.
 \label{eq:fourier-gaussian}
\end{equation}
Equations~\eqref{eq:frel-general}, \eqref{eq:fr-general},
\eqref{eq:relative-gaussian}, and \eqref{eq:recoil-coordinate-gaussian}
then give
\begin{align}
 |F_{\mathrm{rel}}^{(X)}(\bm p)|^2
 &=
 8\pi^{3/2}b_{\mathrm{rel},X}^{3}
 e^{-b_{\mathrm{rel},X}^{2}p^2},
 \label{eq:frel-gaussian}\\
 |F_R(\bm q)|^2
 &=
 8\pi^{3/2}a_{\Phi}^{3}
 e^{-a_{\Phi}^{2}q^2}.
 \label{eq:fr-gaussian}
\end{align}
Consequently, the explicit spatial form factor is
\begin{align}
 |F_X(\bm p,\bm q)|^2
 ={}&
 |\mathcal C_X|^2
 \left(8\pi^{3/2}b_{\mathrm{rel},X}^{3}\right)
 \left(8\pi^{3/2}a_{\Phi}^{3}\right)
 \nonumber\\
 &\times
 e^{-b_{\mathrm{rel},X}^{2}p^2}
 e^{-a_{\Phi}^{2}q^2}.
 \label{eq:explicit-spatial-form-factor}
\end{align}

\subsection{Spin-isospin projection}
\label{subsec:spin-isospin}

The spin-isospin coefficient in
Eq.~\eqref{eq:amplitude-factorization} is included before the phase-space
integration.  The label \(\lambda\) specifies a physical final charge state
and the unobserved spin projections, and
\(\mathcal P_X^{SI}\) is the corresponding incoherent sum defined in
Eq.~\eqref{eq:psi-definition}.

For \(N+{}^{3}N\), the normalized \(T=0\) and \(S=0\) states are
\begin{align}
 |T=0,T_z=0\rangle
 &=
 \frac{1}{\sqrt2}
 \left(|p\,t\rangle-|n\,\HeThree\rangle\right),
 \label{eq:tzero-state}\\
 |S=0,M_S=0\rangle
 &=
 \frac{1}{\sqrt2}
 \left(|\uparrow\downarrow\rangle
 -|\downarrow\uparrow\rangle\right).
 \label{eq:szero-n3n}
\end{align}
For a fixed charge and one of the two nonzero spin product states,
\(|c_{X\lambda}^{SI}|^2=1/4\).  Summing the two orthogonal final spin
configurations gives
\begin{equation}
 \mathcal P_{p+t}^{SI}
 =
 \mathcal P_{n+\HeThree}^{SI}
 =
 \frac14+\frac14=\frac12.
 \label{eq:n3n-spin-isospin}
\end{equation}

Each deuteron has \(S_d=1\) and \(T_d=0\).  For relative orbital
angular momentum \(L=0\), conservation of the initial \(J=0\) selects
\begin{equation}
 \begin{aligned}
 |S=0,M_S=0\rangle_{dd}
 =\frac{1}{\sqrt3}\big(
 &|+1,-1\rangle-|0,0\rangle
 \\
 &+|-1,+1\rangle\big).
 \end{aligned}
 \label{eq:dd-singlet}
\end{equation}
The isospin coupling is unique, and the polarization sum gives
\begin{equation}
 \mathcal P_{d+d}^{SI}
 =
 \sum_{m_1,m_2}
 |\langle 1m_1,1m_2|00\rangle|^2
 =
 \frac13+\frac13+\frac13=1.
 \label{eq:dd-spin-isospin}
\end{equation}
  The separate identical-particle
factor for the two deuterons is included in the phase space below.

It is useful to define the spin-isospin-weighted spatial overlap
\begin{align}
 \mathcal W_X(\bm p,\bm q)
 \equiv{}&
 \sum_{\lambda}
 |c_{X\lambda}^{SI}F_X(\bm p,\bm q)|^2
 \nonumber\\
 ={}&
 \mathcal P_X^{SI}|\mathcal C_X|^2
 \left(8\pi^{3/2}b_{\mathrm{rel},X}^{3}\right)
 \left(8\pi^{3/2}a_{\Phi}^{3}\right)
 \nonumber\\
 &\times
 e^{-b_{\mathrm{rel},X}^{2}p^{2}}
 e^{-a_{\Phi}^{2}q^{2}}.
 \label{eq:spin-summed-overlap}
\end{align}
The complete spin-summed squared transition amplitude is then
\(\sum_{\lambda}|\mathcal M_{X\lambda}|^2
=|\Mcal|^2\mathcal W_X\), in agreement with
Eq.~\eqref{eq:general-spin-summed-amplitude}.

\subsection{Three-body kinematics and phase-space integral}
\label{subsec:phase-space}

Define the \(H\) binding below the \(\Lambda\Lambda\) threshold by
\begin{equation}
 B_{\Lambda\Lambda}^{H}=2m_{\Lambda}-m_H
 \label{eq:H-binding}
\end{equation}
and the measured double-\(\Lambda\) binding of the hypernucleus by
\(B_{\Lambda\Lambda}^{(6)}\).  The intact-$\alpha$ energy release is
\begin{equation}
 Q_{\alpha}
 =
 B_{\Lambda\Lambda}^{H}-B_{\Lambda\Lambda}^{(6)}.
 \label{eq:q-alpha}
\end{equation}
For a breakup partition \(a+b\), let
\begin{equation}
 S_{ab}=m_a+m_b-m_{\alpha}
 \label{eq:breakup-threshold}
\end{equation}
be the alpha breakup threshold.  The three-body energy release is then
\begin{equation}
 Q_X=Q_{\alpha}-S_{ab}.
 \label{eq:q-x}
\end{equation}

The Jacobi reduced masses are
\begin{equation}
 \mu_{ab}=\frac{m_am_b}{m_a+m_b},
 \qquad
 \mu_{H,(ab)}
 =\frac{m_H(m_a+m_b)}{m_H+m_a+m_b}.
 \label{eq:reduced-masses}
\end{equation}
In the nonrelativistic approximation,
\begin{equation}
 Q_X
 =
 \frac{(\hbarc)^2p^2}{2\mu_{ab}}
 +
 \frac{(\hbarc)^2q^2}{2\mu_{H,(ab)}}.
 \label{eq:jacobi-energy}
\end{equation}

The phase-space-overlap factor in
Eq.~\eqref{eq:breakup-factorization} is
\begin{align}
 D_X={}&
 \eta_X
 \int\frac{d^3p}{(2\pi)^3}\frac{d^3q}{(2\pi)^3}
 \delta\!\left(
 Q_X-\frac{(\hbarc)^2p^2}{2\mu_{ab}}
 -\frac{(\hbarc)^2q^2}{2\mu_{H,(ab)}}
 \right)
 \nonumber\\
 &\times\mathcal W_X(\bm p,\bm q),
 \label{eq:dx-sixdim}
\end{align}
where
\begin{equation}
 \eta_{p+t}=\eta_{n+\HeThree}=1,
 \qquad
 \eta_{d+d}=\frac{1}{2!}.
 \label{eq:eta-values}
\end{equation}
The \(d+d\) factor prevents double counting of identical final deuterons.

Assigning energy \(E\) to the \(a\)-\(b\) Jacobi motion gives
\begin{equation}
 \begin{aligned}
 p(E)&=\frac{\sqrt{2\mu_{ab}E}}{\hbarc},
 \\
 q(Q_X-E)&=
 \frac{\sqrt{2\mu_{H,(ab)}(Q_X-E)}}{\hbarc}.
 \end{aligned}
 \label{eq:pq-energy}
\end{equation}
After the angular integrations and the energy-conserving delta function,
Eq.~\eqref{eq:dx-sixdim} becomes
\begin{align}
 D_X={}&
 \eta_X\mathcal P_X^{SI}|\mathcal C_X|^2
 \frac{(4\pi)^2}{(2\pi)^6}
 \frac{2(\mu_{ab}\mu_{H,(ab)})^{3/2}}{(\hbarc)^6}
 \nonumber\\
 &\times
 \left(8\pi^{3/2}b_{\mathrm{rel},X}^{3}\right)
 \left(8\pi^{3/2}a_{\Phi}^{3}\right)
 \mathcal N_X,
 \label{eq:dx-one-d}
\end{align}
with
\begin{equation}
 \mathcal N_X=
 \int_0^{Q_X}dE\,
 \sqrt{E(Q_X-E)}
 e^{-A_XE-B_X(Q_X-E)}.
 \label{eq:nx-integral}
\end{equation}
The inverse-energy coefficients are
\begin{equation}
 A_X=\frac{2b_{\mathrm{rel},X}^{2}\mu_{ab}}{(\hbarc)^2},
 \qquad
 B_X=\frac{2a_{\Phi}^{2}\mu_{H,(ab)}}{(\hbarc)^2}.
 \label{eq:ab-coefficients}
\end{equation}

The integral in Eq.~\eqref{eq:nx-integral} also has a closed form.  With
\begin{equation}
 z_X=\frac{(B_X-A_X)Q_X}{2},
 \label{eq:z-definition}
\end{equation}
one obtains
\begin{equation}
 \mathcal N_X
 =
 \frac{\pi Q_X^2}{4}
 e^{-(A_X+B_X)Q_X/2}
 \frac{I_1(z_X)}{z_X},
 \label{eq:nx-analytic}
\end{equation}
where \(I_1\) is the modified Bessel function of the first kind and
\(I_1(z)/z\rightarrow1/2\) as \(z\rightarrow0\).

Then we can take an average over the three-body phase-space to see what fraction of the available three-body phase-space survives the Gaussian form-factor suppression.
The meaning of the phase-space average can be stated explicitly in terms
of the energy-sharing variable $E$. In Eq.~\eqref{eq:pq-energy}, $E$ is
the kinetic energy of the relative $a$--$b$ motion, whereas $Q_X-E$ is
the kinetic energy of the $H$ relative to the center of mass of the
$a+b$ pair. After the angular integrations, the nonrelativistic
three-body phase-space measure is proportional to
\begin{equation}
 dE\,\sqrt{E(Q_X-E)}.
 \label{eq:energy-sharing-measure}
\end{equation}

For any function $f(E)$, we define its normalized three-body phase-space average as
\begin{align}
 \left\langle f\right\rangle_{\Phi_3}
 &\equiv
 \frac{
  \displaystyle\int_0^{Q_X}dE\,
  \sqrt{E(Q_X-E)}\,f(E)
 }{
  \displaystyle\int_0^{Q_X}dE\,
  \sqrt{E(Q_X-E)}
 }
 \nonumber\\
 &=
 \frac{8}{\pi Q_X^2}
 \int_0^{Q_X}dE\,
 \sqrt{E(Q_X-E)}\,f(E).
 \label{eq:phase-space-average}
\end{align}

The denominator is therefore the unweighted energy-sharing integral
obtained by setting the momentum-dependent Gaussian equal to unity.
For the Gaussian form factor used in the present calculation, the
function entering the phase-space average is
\begin{equation}
 f_X(E)
 =
 \exp\!\left[-A_XE-B_X(Q_X-E)\right].
 \label{eq:gaussian-energy-weight}
\end{equation}
The phase-space-averaged Gaussian shape factor is consequently
\begin{align}
 \mathcal S_X
 &\equiv
 \left\langle f_X\right\rangle_{\Phi_3}
 \nonumber\\
 &=
 \frac{8\mathcal N_X}{\pi Q_X^2}
 \nonumber\\
 &=
 2e^{-(A_X+B_X)Q_X/2}
 \frac{I_1(z_X)}{z_X}.
 \label{eq:average-shape}
\end{align}
Thus, $\mathcal S_X$ is the ratio of the Gaussian-weighted
energy-sharing integral to the corresponding unweighted integral. It
describes only the suppression generated by the momentum dependence of
the two Gaussian Fourier factors.

Since the Fourier-normalization volumes are already kept explicitly in
Eq.~\eqref{eq:dx-one-d}, the remaining dimensionless suppression weight
is
\begin{equation}
 \overline{\mathcal W}_X
 =
 \mathcal P_X^{SI}
 |\mathcal C_X|^2
 \mathcal S_X.
 \label{eq:average-full}
\end{equation}
Here $|\mathcal C_X|^2$ is the internal cluster-overlap factor, while
$\mathcal P_X^{SI}$ is the spin-isospin factor obtained after summing
over the unobserved final spin states. The Fourier-normalization volumes
and the identical-particle factor $\eta_X$ are not included in
$\overline{\mathcal W}_X$; they remain explicit in the expression for
$D_X$.

\section{Results}
\label{sec:results}

\subsection{Numerical inputs}
\label{subsec:inputs}

Table~\ref{tab:inputs} collects all masses, radii, constants, and Gal
normalization data used in the central calculation.  Nuclear masses are
taken from standard mass evaluations, while \(m_{\Lambda}\) is taken from
the Particle Data Group \cite{Wang2021,Navas2024}.  The light-nuclear
charge radii are from the Angeli--Marinova compilation
\cite{Angeli2013}, and the values of \(\hbar\) and \(\hbarc\) follow the
CODATA convention \cite{Tiesinga2021}.

\begin{table*}[t]
\centering
\caption{\label{tab:inputs}
Numerical inputs.  The \(H\)-mass entry is the rounded
\(\Lambda+n\) threshold benchmark.  To remain exactly consistent with
Gal's tabulated calculation, the energy release is determined from
\(B_{\Lambda\Lambda}^{H}=176~\mathrm{MeV}\), while
\(m_H=2055~\mathrm{MeV}\) is used in reduced masses.}
%\small
\begin{tabular}{@{}lcl@{\qquad}lcl@{}}
\toprule
Quantity & Value & Unit & Quantity & Value & Unit\\
\midrule
\(m_H\) & \(2055.000\) & \(\mathrm{MeV}\) &
\(B_{\Lambda\Lambda}^{H}\) & \(176.000\) & \(\mathrm{MeV}\)\\
\(m_{\Lambda}\) & \(1115.683\) & \(\mathrm{MeV}\) &
\(B_{\Lambda\Lambda}^{(6)}\) & \(6.910\) & \(\mathrm{MeV}\)\\
\(m_p\) & \(938.2720813\) & \(\mathrm{MeV}\) &
\(m_n\) & \(939.5654133\) & \(\mathrm{MeV}\)\\
\(m_d\) & \(1875.6129426\) & \(\mathrm{MeV}\) &
\(m_t\) & \(2808.9211330\) & \(\mathrm{MeV}\)\\
\(m_{{}^{3}\mathrm{He}}\) & \(2808.3916074\) & \(\mathrm{MeV}\) &
\(m_{\alpha}\) & \(3727.3794066\) & \(\mathrm{MeV}\)\\
\(r_{\mathrm{ch}}(d)\) & \(2.1421\) & \(\mathrm{fm}\) &
\(r_{\mathrm{ch}}(t)\) & \(1.7591\) & \(\mathrm{fm}\)\\
\(r_{\mathrm{ch}}({}^{3}\mathrm{He})\) & \(1.9661\) & \(\mathrm{fm}\) &
\(r_{\mathrm{ch}}(\HeFour)\) & \(1.6755\) & \(\mathrm{fm}\)\\
\(a_{\Phi}\) & \(1.490\) & \(\mathrm{fm}\) &
\(\hbarc\) & \(197.3269804\) & \(\mathrm{MeV\,fm}\)\\
\(\hbar\) & \(6.582119569\times10^{-16}\) & \(\mathrm{eV\,s}\) &
\(k_{\alpha}^{\mathrm{Gal}}\) & \(3.393\) & \(\mathrm{fm}^{-1}\)\\
\(I_{\alpha}^{\mathrm{Gal}}\) & \(1.521\times10^{-8}\) & \(\mathrm{fm}^{3}\) &
\(\Gamma_{\alpha}^{\mathrm{Gal}}\) & \(1.550\times10^{-7}\) & \(\mathrm{eV}\)\\
\bottomrule
\end{tabular}
\end{table*}

The benchmark energy release in the intact-$\alpha$ channel is
\begin{equation}
 Q_{\alpha}=176.000-6.910=169.090~\mathrm{MeV}.
 \label{eq:qalpha-number}
\end{equation}
Although the table lists \(m_H\) to the nearest MeV, using the binding
energy in Eq.~\eqref{eq:qalpha-number} preserves Gal's benchmark exactly.
The rounding changes the reduced masses by less than \(0.02\%\).

\subsection{Gaussian widths and internal overlaps}
\label{subsec:numerical-widths}

Equation~\eqref{eq:bn-alpha} gives
\begin{equation}
 b_N=1.57968~\mathrm{fm}.
 \label{eq:bn-number}
\end{equation}
The corresponding separation widths are
\begin{equation}
 \begin{aligned}
 b_{\mathrm{rel}}^{(1+3)}
 &=b_N\sqrt{\frac43}=1.82405~\mathrm{fm},
 \\
 b_{\mathrm{rel}}^{(2+2)}
 &=b_N=1.57968~\mathrm{fm}.
 \end{aligned}
 \label{eq:brel-numbers}
\end{equation}

Table~\ref{tab:cluster-widths} shows the intrinsic widths and Gaussian
overlaps of the final clusters.  For \(p+t\), the proton has no intrinsic
coordinate and
\(|\mathcal C_{p+t}|^2=\mathcal O_t^2\).  Similarly,
\(|\mathcal C_{n+\HeThree}|^2=\mathcal O_{{}^{3}\mathrm{He}}^2\), while
the two deuteron overlaps give
\(|\mathcal C_{d+d}|^2=\mathcal O_d^4\).

\begin{table}[t]
\centering
\caption{\label{tab:cluster-widths}
Intrinsic Gaussian widths and overlap amplitudes. The widths
$b_c$ are inferred from the experimental charge radii using
Eq.~\eqref{eq:b-from-radius}; for the alpha particle,
$b_{\alpha}=b_N$. The overlap amplitudes are then obtained from
Eq.~\eqref{eq:cluster-overlap}.}
%\scriptsize
\begin{tabular}{@{}lcccc@{}}
\toprule
Cluster & \(A_c\) & \(r_{\mathrm{ch}}\) [fm] &
\(b_c\) [fm] & \(\mathcal O_c\)\\
\midrule
\(t\) & \(3\) & \(1.7591\) & \(1.75910\) & \(0.98282\)\\
\({}^{3}\mathrm{He}\) & \(3\) & \(1.9661\) & \(1.96610\) & \(0.93122\)\\
\(d\) & \(2\) & \(2.1421\) & \(2.47348\) & \(0.86415\)\\
\(\alpha\) & \(4\) & \(1.6755\) & \(1.57968\) & ---\\
\bottomrule
\end{tabular}
\end{table}

The channel factors entering the rate are
\begin{align}
 |\mathcal C_{p+t}|^2&=0.96594, &
 \mathcal P_{p+t}^{SI}&=\frac12, \nonumber\\
 |\mathcal C_{n+\HeThree}|^2&=0.86716, &
 \mathcal P_{n+\HeThree}^{SI}&=\frac12, \nonumber\\
 |\mathcal C_{d+d}|^2&=0.55765, &
 \mathcal P_{d+d}^{SI}&=1.
 \label{eq:channel-factors-numerical}
\end{align}

\subsection{Kinematics}
\label{subsec:numerical-kinematics}

The thresholds and energy releases obtained from
Eqs.~\eqref{eq:breakup-threshold} and \eqref{eq:q-x} are
\begin{align}
 S_{p+t}&=19.81381~\mathrm{MeV}, &
 Q_{p+t}&=149.27619~\mathrm{MeV}, \nonumber\\
 S_{n+\HeThree}&=20.57761~\mathrm{MeV}, &
 Q_{n+\HeThree}&=148.51239~\mathrm{MeV}, \nonumber\\
 S_{d+d}&=23.84648~\mathrm{MeV}, &
 Q_{d+d}&=145.24352~\mathrm{MeV}.
 \label{eq:q-values}
\end{align}
All three breakup channels are therefore widely open.  The reduced masses
and exponential coefficients used in the integral are listed in
Table~\ref{tab:kinematics}.

\begin{table*}[t]
\centering
\caption{\label{tab:kinematics}
Kinematic and Gaussian coefficients for the three breakup channels.
The coefficients \(A_X\) and \(B_X\) are defined in
Eq.~\eqref{eq:ab-coefficients}; \(z_X\) is dimensionless.}
%\small
\begin{tabular}{@{}lrrrrrrr@{}}
\toprule
Channel &
\(S_{ab}\) [MeV] &
\(Q_X\) [MeV] &
\(\mu_{ab}\) [MeV] &
\(\mu_{H,(ab)}\) [MeV] &
\(A_X\) [MeV\(^{-1}\)] &
\(B_X\) [MeV\(^{-1}\)] &
\(z_X\)\\
\midrule
\(H+p+t\) &
\(19.81381\) & \(149.27619\) & \(703.33504\) & \(1327.16746\) &
\(0.120197\) & \(0.151341\) & \(2.32448\)\\
\(H+n+\HeThree\) &
\(20.57761\) & \(148.51239\) & \(704.02825\) & \(1327.26326\) &
\(0.120316\) & \(0.151351\) & \(2.30460\)\\
\(H+d+d\) &
\(23.84648\) & \(145.24352\) & \(937.80647\) & \(1327.67297\) &
\(0.120201\) & \(0.151398\) & \(2.26562\)\\
\bottomrule
\end{tabular}
\end{table*}

\subsection{Integrated form-factor weights, widths, and lifetime}
\label{subsec:numerical-rates}

Using Gal's tabulated \(k_{\alpha}\) and
\(I(k_{\alpha},a_{\Phi})\) in Eq.~\eqref{eq:gal-D} gives
\begin{equation}
 \begin{aligned}
 \mu_{H\alpha}&=1324.67349~\mathrm{MeV},
 \\
 D_{\alpha}^{\mathrm{Gal}}
 &=4.44723\times10^{-11}~\mathrm{MeV}^{-1}.
 \end{aligned}
 \label{eq:dalpha-number}
\end{equation}
The one-dimensional phase-space integrals in
Eq.~\eqref{eq:nx-integral} were evaluated both numerically and with
Eq.~\eqref{eq:nx-analytic}; the two evaluations agree to better than
\(10^{-11}\) in relative precision.

Table~\ref{tab:rates} gives the shape averages, the full averages after
inserting the internal overlap and spin-isospin factors, the phase-space-overlap
factors \(D_X\), the ratios \(R_X\), and the partial widths.  The
identical-deuteron factor \(\eta_{d+d}=1/2\) is already included in the
listed \(D_{d+d}\).

\begin{table*}[t]
\centering
\caption{\label{tab:rates}
Integrated results for the integral \(\mathcal N_X\), the shape averages \(\mathcal S_X\), the full averages after
inserting the internal and spin-isospin factors \(\overline{\mathcal W}_X\), the phase-space-overlap
factors \(D_X\), and the ratios \(R_X\).}
%\scriptsize
\begin{tabular}{@{}lrrrrrr@{}}
\toprule
Channel &
\(\mathcal N_X\) [$\mathrm{MeV}^2$] &
\(\mathcal S_X\) &
\(\overline{\mathcal W}_X\) &
\(D_X\) [$\mathrm{MeV}^{-1}$] &
\(R_X\) &
\(\Gamma_X\) [eV]\\
\midrule
\(H+p+t\) &
\(2.54887\times10^{-5}\) &
\(2.91278\times10^{-9}\) &
\(1.40678\times10^{-9}\) &
\(3.84550\times10^{-8}\) &
\(8.64696\times10^{2}\) &
\(1.34028\times10^{-4}\)\\
\(H+n+\HeThree\) &
\(2.74537\times10^{-5}\) &
\(3.16968\times10^{-9}\) &
\(1.37431\times10^{-9}\) &
\(3.72430\times10^{-8}\) &
\(8.37442\times10^{2}\) &
\(1.29804\times10^{-4}\)\\
\(H+d+d\) &
\(4.03817\times10^{-5}\) &
\(4.87452\times10^{-9}\) &
\(2.71830\times10^{-9}\) &
\(3.51942\times10^{-8}\) &
\(7.91375\times10^{2}\) &
\(1.22663\times10^{-4}\)\\
\bottomrule
\end{tabular}
\end{table*}

Among the three alpha-breakup modes, the \(H+p+t\) channel provides the largest individual contribution, with \(\Gamma_{p+t}=1.34028\times10^{-4}\,\mathrm{eV}\), followed closely by \(\Gamma_{n+{}^{3}\mathrm{He}}=1.29804\times10^{-4}\,\mathrm{eV}\) and \(\Gamma_{d+d}=1.22663\times10^{-4}\,\mathrm{eV}\). These values correspond to \(34.68\%\), \(33.58\%\), and \(31.74\%\), respectively, of the total breakup width. Therefore, the \(p+t\) channel is the leading contribution, but no single breakup channel dominates the result. Since the short-distance factor \(\lvert M_0\rvert^2\) is common to all channels, their relative importance is determined by the complete phase-space-overlap factor \(D_X\), for which \(D_{p+t}=3.84550\times10^{-8}\,\mathrm{MeV}^{-1}\), \(D_{n+{}^{3}\mathrm{He}}=3.72430\times10^{-8}\,\mathrm{MeV}^{-1}\), and \(D_{d+d}=3.51942\times10^{-8}\,\mathrm{MeV}^{-1}\). The modest preference for \(p+t\) arises mainly from its larger internal overlap, \(\lvert C_{p+t}\rvert^2=0.96594\), and its slightly larger energy release. For the \(d+d\) channel, the smaller internal overlap and lower energy release are partially compensated by its smaller cluster-separation width, larger reduced mass, and larger phase-space integral, thereby keeping all three partial widths comparable.

The total enhancement relative to the coherent intact-$\alpha$ channel is
\begin{equation}
 R_{\mathrm{br}}
 =
 R_{p+t}+R_{n+\HeThree}+R_{d+d}
 =
 2.49351\times10^{3}.
 \label{eq:rbreak-number}
\end{equation}
The breakup width reads
\begin{equation}
 \Gamma_{\mathrm{br}}
 = \Gamma_{p+t}
+\Gamma_{n+{}^{3}\mathrm{He}}
+\Gamma_{d+d}=
 3.86495\times10^{-4}~\mathrm{eV}.
 \label{eq:gammabreak-number}
\end{equation}
Adding the intact-$\alpha$ contribution, \(\Gamma_{\alpha}=1.55\times10^{-7}~\mathrm{eV}\), gives
\begin{equation}
 \Gamma_{\mathrm{inc}}
 =
 \Gamma_{\alpha}+\Gamma_{\mathrm{br}}
 =
 3.86650\times10^{-4}~\mathrm{eV},
 \label{eq:gammainc-number}
\end{equation}
and hence
\begin{equation}
 \tau_{\mathrm{inc}}
 =
 \frac{\hbar}{\Gamma_{\mathrm{inc}}}
 =
 1.70235\times10^{-12}~\mathrm{s}.
 \label{eq:tauinc-number}
\end{equation}

\subsection{Physical interpretation}
\label{subsec:interpretation}

The magnitude of the calculated breakup width is strongly controlled by
the spatial scales entering the Gaussian form factors. In particular,
the alpha-derived separation widths $b_{\mathrm{rel},X}$ enter both the
Fourier normalization and, more importantly, the exponential
momentum-space suppression. A larger coordinate-space width corresponds
to a narrower momentum-space distribution. Therefore, at the relatively
large Jacobi momenta involved in the present decay, increasing
$b_{\mathrm{rel},X}$ decreases $D_X$ and $\Gamma_X$ and consequently
increases the lifetime. A smaller separation width has the opposite
effect and can produce a substantially larger breakup width.

This dependence is exponential rather than merely algebraic. As a
diagnostic sensitivity test, rescaling the alpha Gaussian width $b_N$,
and hence all separation widths $b_{\mathrm{rel},X}$, by $-10\%$ changes
the inclusive lifetime from the central value
$1.70\times10^{-12}\ {\rm s}$ to approximately
$1.94\times10^{-13}\ {\rm s}$. A corresponding $+10\%$ rescaling gives
approximately $1.30\times10^{-11}\ {\rm s}$. The internal overlap
factors are recomputed consistently in this test, whereas the
final-cluster widths $b_c$ and the recoil width $a_{\Phi}$ are held
fixed. This variation is not an experimental uncertainty in the charge
radius; rather, it illustrates the structural sensitivity associated
with representing the nuclear wave functions by single Gaussians.

A further source of model dependence, not covered by the width-rescaling
test above, concerns the high-momentum behavior of the relative motion
between the $\Lambda\Lambda$ subsystem and the $\alpha$ core. With the
single-Gaussian ansatz adopted for $\Phi_{\Lambda\Lambda}(R)$,
Eq.~\eqref{eq:fr-gaussian} imposes the exponential behavior
\[
    |F_R(q)|^2 \propto \exp(-a_\Phi^2 q^2).
\]
At the relatively large momentum transfers relevant to the formation of a
deeply bound $H$ dibaryon, however, the overlap probes the short-distance
structure of the initial hypernuclear wave function. Short-range
correlations and other non-Gaussian components of a realistic wave
function can generate a more slowly decreasing high-momentum tail than
that produced by a single Gaussian
\cite{Lonardoni2017, Hen:2016kwk}. Such a tail would generally enhance the
momentum-space overlap, tending to increase the strong-decay width and
therefore shorten the corresponding lifetime. Since the decay rate
depends quadratically on the transition amplitude, even a moderate
enhancement of the high-momentum overlap could potentially lead to an
order-of-magnitude change in the predicted width or lifetime. 
%This possible correction is distinct from the short-range correlation in the internal $\Lambda\Lambda$ coordinate $r_{\Lambda\Lambda}$ entering the short-distance amplitude $M_0$ in Eqs.~(8) and~(9) \cite{Gal2024}. 
Since $F_R$ enters both the intact-$\alpha$
channel and the breakup calculation, its effect cannot be
incorporated consistently by modifying only the breakup factor $D_X$
while retaining the original Gal normalization. A quantitative estimate
would require a correlated $\Lambda\Lambda$--$\alpha$ wave function and
a simultaneous reevaluation of the intact-$\alpha$ and breakup
contributions. The lifetimes reported here should therefore be regarded
as central estimates within the adopted Gaussian framework.

The internal cluster widths $b_c$ play a secondary and conceptually
different role. They enter through the overlap factors
$|\mathcal C_X|^2$, which are maximal when the intrinsic width of the
outgoing cluster matches the corresponding alpha-particle width and
decrease as the mismatch grows. Thus, it is specifically the increase
of $b_{\mathrm{rel},X}$, rather than a universal increase of every
Gaussian width, that is responsible for the large increase in the
predicted lifetime.

For the central widths adopted here, the total breakup contribution
still exceeds the Gal-normalized intact-$\alpha$ contribution by
$R_{\mathrm{br}}=2.49\times10^{3}$. Although the larger separation
widths strongly suppress the breakup form factors, the coherent
intact-$\alpha$ overlap at $k_{\alpha}\simeq3.4\ {\rm fm}^{-1}$ is even
more strongly suppressed. In the three-body channels, the available
energy can be shared between the $a$--$b$ relative motion and the recoil
of the $H$ against the $ab$ pair. This additional degree of freedom,
together with the three-body phase-space volume, allows the breakup
channels to remain dominant. The resulting central lifetime,
$1.70\times10^{-12}\ {\rm s}$, lies within approximately two orders of
magnitude of the characteristic hypernuclear weak-decay scale of
$10^{-10}\ {\rm s}$.

A second important source of sensitivity is the assumed $H$-dibaryon
mass. The present calculation uses $m_H\simeq2055\ {\rm MeV}$, close to
the $m_{\Lambda}+m_n$ threshold, whereas the frequently discussed
stable-sexaquark dark-matter window is approximately
$m_H=1860$--$1880\ {\rm MeV}$ \cite{Farrar2022Overview}. For example, decreasing
the mass to $m_H\simeq1870\ {\rm MeV}$ increases the breakup energy
releases from approximately $145$--$149\ {\rm MeV}$ to approximately
$331$--$335\ {\rm MeV}$. Although this increases the available
three-body phase space, it also requires substantially larger Jacobi
momenta. Within the present Gaussian description, the resulting
exponential form-factor suppression is expected to dominate the
polynomial increase of phase space, leading to a smaller strong-decay
width and a longer lifetime.

Finally, spin-isospin projection divides the normalized $T=0$ strength
equally between the $p+t$ and $n+{}^{3}\mathrm{He}$ charge channels.
The $d+d$ channel retains the full normalized $S=0$, $T=0$ strength,
while its phase space contains the separate identical-particle factor
$1/2$.

\section{Comparison with Gal's calculation}
\label{sec:comparison_gal}

 Gal considered the intact-$\alpha$ decay channel Eq.~\eqref{eq:intact_alpha}
and calculated its decay width for five representative values of the
$H$-dibaryon binding energy \cite{Gal2024}. For each value of $B_{\Lambda\Lambda}^{H}$ listed in Table~2 of
Gal's calculation, the corresponding $H$ mass is obtained from Eq.~\eqref{eq:H-binding} where $m_{\Lambda}=1115.683~\mathrm{MeV}$. In the present calculation,
we additionally include the three alpha-breakup channels. The total breakup and
inclusive widths are therefore defined in Eq.~\eqref{eq:gammabreak-number} and \eqref{eq:gammainc-number} respectively,
with the corresponding inclusive lifetime given in Eq.~\eqref{eq:tauinc-number}.
 The resulting comparison is
presented in Table~\ref{tab:gal_comparison}.

\begin{table*}[t]
\centering
\caption{Comparison of Gal's intact-$\alpha$ result with the inclusive
result obtained in the present calculation at the five binding energies
listed in Table~2 of Ref.~\cite{Gal2024}. The corresponding $H$ masses
are obtained from
$m_H=2m_{\Lambda}-B_{\Lambda\Lambda}^{H}$.}
\label{tab:gal_comparison}
%\small
\begin{tabular*}{\textwidth}
{@{\extracolsep{\fill}}ccccccc}
\hline\hline
$B_{\Lambda\Lambda}^{H}$
&
$m_H$
&
$\Gamma_{\alpha}^{\mathrm{Gal}}$
&
$\tau_{\alpha}^{\mathrm{Gal}}$
&
$\Gamma_{\mathrm{br}}$
&
$\Gamma_{\mathrm{inc}}$
&
$\tau_{\mathrm{inc}}$
\\
(MeV)
&
(MeV)
&
(eV)
&
(s)
&
(eV)
&
(eV)
&
(s)
\\
\hline
100
&
2131.366
&
$7.820\times10^{-3}$
&
$8.410\times10^{-14}$
&
$1.661\times10^{0}$
&
$1.669\times10^{0}$
&
$3.943\times10^{-16}$
\\
\textbf{176}
&
\textbf{2055.366}
&
$\bm{1.550\times10^{-7}}$
&
$\bm{4.245\times10^{-9}}$
&
$\bm{3.863\times10^{-4}}$
&
$\bm{3.864\times10^{-4}}$
&
$\bm{1.703\times10^{-12}}$
\\
200
&
2031.366
&
$5.010\times10^{-9}$
&
$1.315\times10^{-7}$
&
$2.462\times10^{-5}$
&
$2.463\times10^{-5}$
&
$2.673\times10^{-11}$
\\
300
&
1931.366
&
$6.790\times10^{-15}$
&
$9.700\times10^{-2}$
&
$2.238\times10^{-10}$
&
$2.238\times10^{-10}$
&
$2.941\times10^{-6}$
\\
400
&
1831.366
&
$2.436\times10^{-20}$
&
$2.703\times10^{4}$
&
$1.921\times10^{-15}$
&
$1.921\times10^{-15}$
&
$3.426\times10^{-1}$
\\
\hline\hline
\end{tabular*}
\end{table*}

Table~\ref{tab:gal_comparison} demonstrates that the combined
contribution of the three breakup channels is larger than the
intact-$\alpha$ contribution at all five benchmark points. The effect is
particularly significant for
$B_{\Lambda\Lambda}^{H}=176~\mathrm{MeV}$, corresponding to
$m_H=2055.366~\mathrm{MeV}$. At this point, Gal's intact-$\alpha$
calculation gives
$\tau_{\alpha}^{\mathrm{Gal}}=4.245\times10^{-9}~\mathrm{s}$ \cite{Gal2024},
whereas inclusion of the breakup channels reduces the lifetime to
$\tau_{\mathrm{inc}}=1.703\times10^{-12}~\mathrm{s}$. Therefore,
although the intact-$\alpha$ lifetime is longer than the characteristic
weak-decay timescale of a double-Lambda hypernucleus, the inclusive
lifetime at this mass is considerably shorter.

\begin{figure}[t]
\centering
\includegraphics[width=0.48\textwidth]{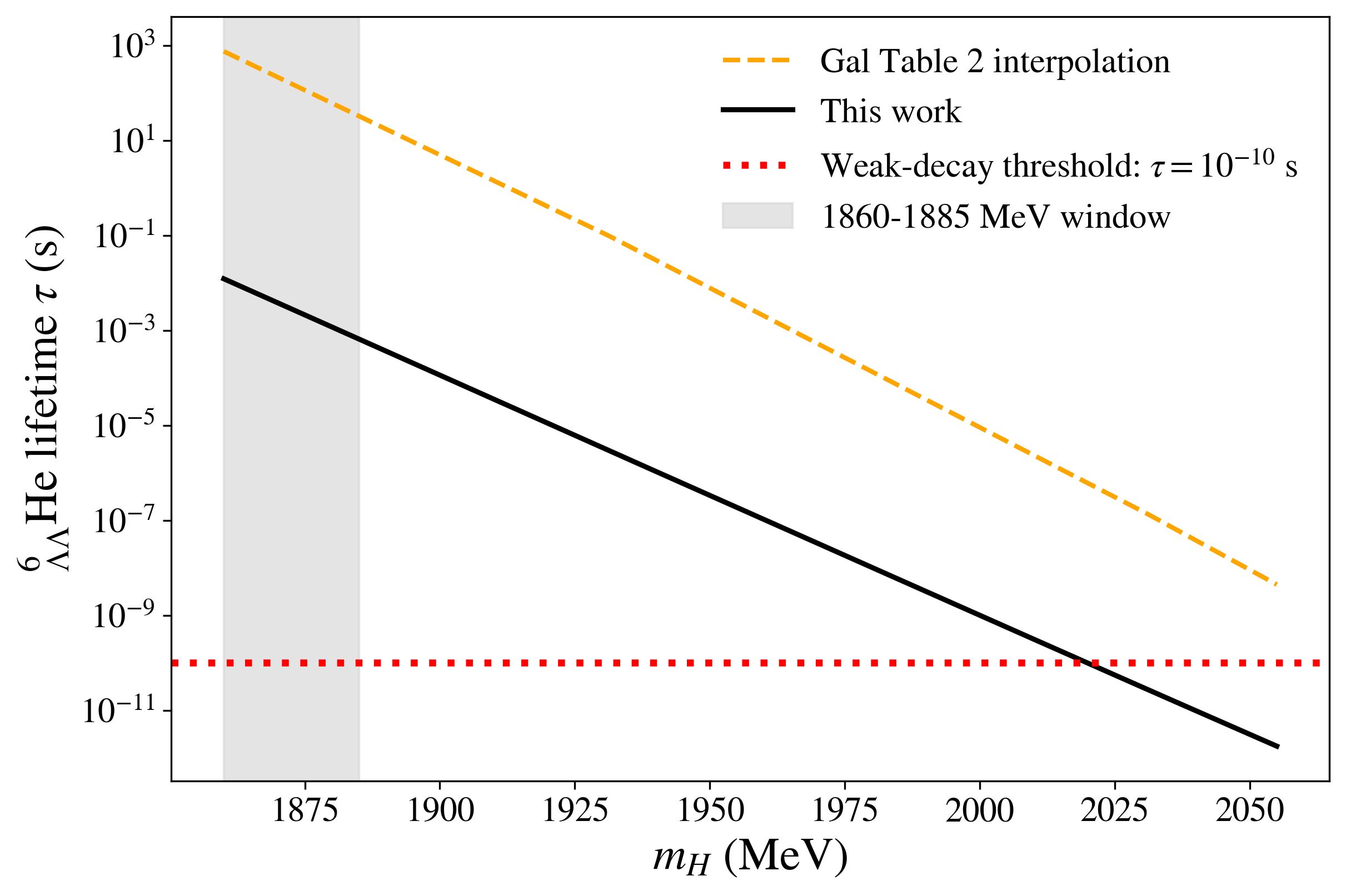}
\caption{The intact-$\alpha$ lifetime (orange dashed curve) is obtained by interpolating the
benchmark results of Ref.~\cite{Gal2024}, and the inclusive lifetime is obtained
after including the three alpha-breakup channels (black solid curve)
as functions of the $H$-dibaryon mass. The horizontal red dotted line denotes
the characteristic weak-decay timescale of the parent double-$\Lambda$
hypernucleus, $\tau_{\mathrm{weak}}=10^{-10}~\mathrm{s}$ \cite{Gal2024, Hiyama2018}. We highlight the representative dark-matter-motivated interval, $1865\leq m_H\leq1885~\mathrm{MeV}$, by the shaded gray region \cite{farrar2022,Moore:2024mot,Moore:2025zyg, Buccella:2020mxi}.
}
\label{fig:lifetime_mass_scan}
\end{figure}

To investigate the mass dependence, we extend the calculation over a
continuous range of $H$ masses. Between the five binding energies
listed by Gal \cite{Gal2024}, $k_H$ is interpolated linearly, while the overlap
integral and the intact-$\alpha$ decay width are interpolated
logarithmically. The resulting intact-$\alpha$ and inclusive lifetimes are
shown in Fig.~\ref{fig:lifetime_mass_scan}. At each selected value of $m_H$, all breakup quantities are calculated within our model while the intact-$\alpha$ quantities taken from Gal's tabulated
calculation are interpolated between the published benchmark points.

The inclusive lifetime reaches the characteristic hypernuclear
weak-decay timescale at $m_H^{\mathrm{cross}}
\simeq 2.020~\mathrm{GeV}$ corresponding to
$B_{\Lambda\Lambda}^{H}
\simeq 211.42~\mathrm{MeV}$.
Within the present calculation, the inclusive lifetime satisfies
\begin{equation}
\tau_{\mathrm{inc}}>10^{-10}~\mathrm{s}
\qquad
\text{for}
\qquad
m_H<2.020~\mathrm{GeV}.
\end{equation}
Therefore, below approximately $2020~\mathrm{MeV}$, the strong
conversion of
${}^{6}_{\Lambda\Lambda}\mathrm{He}$ into final states containing an
$H$ dibaryon is slower than the characteristic weak decay of the
parent hypernucleus.

In the dark matter-motivated mass interval
$1865\leq m_H\leq1885~\mathrm{MeV}$ \cite{farrar2022,Moore:2024mot,Moore:2025zyg, Buccella:2020mxi}, the inclusive lifetime varies
from
\begin{equation}
\tau_{\mathrm{inc}}
=
6.80\times10^{-3}~\mathrm{s}
\end{equation}
at $m_H=1865~\mathrm{MeV}$ to
\begin{equation}
\tau_{\mathrm{inc}}
=
6.59\times10^{-4}~\mathrm{s}
\end{equation}
at $m_H=1885~\mathrm{MeV}$. These lifetimes are approximately
$10^{6}$--$10^{8}$ times longer than the characteristic
hypernuclear weak-decay timescale. Consequently, the observation of
weakly decaying double-Lambda hypernuclei does not exclude a deeply
bound $H$ dibaryon in this low-mass interval.

This conclusion also applies to the mass region favored in our
previous astrophysical analysis \cite{Shahrbaf:2024gdm}. In that work, the
Bayesian analysis favored sexaquark masses below approximately
$1935~\mathrm{MeV}$. At the upper boundary of this region, the present
calculation gives
\begin{equation}
\tau_{\mathrm{inc}}
\left(m_H=1935~\mathrm{MeV}\right)
=
1.93\times10^{-6}~\mathrm{s},
\end{equation}
which is approximately $1.9\times10^{4}$ times longer than
$10^{-10}~\mathrm{s}$. Since the inclusive lifetime increases as the
$H$ mass decreases, the full mass region below $1935~\mathrm{MeV}$
considered here satisfies
$\tau_{\mathrm{inc}}\gg10^{-10}~\mathrm{s}$.

We therefore conclude that, although the alpha-breakup channels
substantially decrease the lifetime relative to the intact-$\alpha$
calculation, they do not exclude a deeply bound $H$ dibaryon in the
low-mass region relevant to sexaquark dark matter or in the mass range
favored by our previous neutron star analysis. Within the assumptions
of the present calculation, these mass ranges remain compatible with
the existing double-Lambda hypernuclear decay data. 
%This conclusion refers specifically to the competition between strong $H$ production and weak decay of the parent hypernucleus; the weak lifetime of the $H$ dibaryon itself is a separate question.

\section{Conclusions}
\label{sec:conclusions}

We have quantified the contribution of $\alpha$-core breakup to the strong
conversion of \(\HeLL\) into a deeply bound \(H\) dibaryon.  The calculation
extends the coherent \(H+\HeFour\) channel considered by Gal by explicitly
including the three-body final states \(H+p+\tritium\), \(H+n+\HeThree\), and
\(H+d+d\).  These channels were treated within a translationally invariant
Gaussian cluster framework, with spin-isospin recoupling included at the
amplitude level and the corresponding two-Jacobi-momentum phase spaces
integrated numerically.

At the benchmark \(B_{\Lambda\Lambda}^{H}=176~\mathrm{MeV}\), corresponding
to \(m_H\simeq2055~\mathrm{MeV}\), the summed breakup width exceeds the
Gal-normalized intact-\(\alpha\) width by
\(R_{\mathrm{br}}=2.49\times10^{3}\).  The inclusive result is $\Gamma_{\mathrm{inc}}=3.87\times10^{-4}~\mathrm{eV}$, and $\tau_{\mathrm{inc}}=1.70\times10^{-12}~\mathrm{s}$
compared with \(\tau_{\alpha}=4.25\times10^{-9}~\mathrm{s}\) when only the
coherent intact-\(\alpha\) channel is retained.  Core breakup therefore
shortens the strong-conversion lifetime by more than three orders of
magnitude at this benchmark.

The enhancement has a clear kinematic origin.  In the coherent two-body
channel, the intact $\alpha$ particle must absorb the full recoil momentum,
and the corresponding high-momentum overlap is strongly suppressed.  In
the three-body breakup channels, the available energy can instead be
distributed between the relative motion of the two nuclear fragments and
the recoil of their center of mass against the \(H\).  This additional
degree of freedom, together with the larger three-body phase-space volume,
allows the breakup channels to dominate even though their individual
cluster form factors are themselves suppressed.  The coherent
\(H+\alpha\) channel therefore does not, in general, saturate the inclusive
strong width.

The mass-dependent calculation reveals a competition between the increasing
energy release and the decreasing high-momentum overlap.  As \(m_H\)
decreases, the available three-body phase space grows, but progressively
larger Jacobi momenta are required.  Within the Gaussian description, the
resulting exponential suppression eventually dominates, causing the
inclusive lifetime to increase rapidly toward lower \(H\) masses.  The
calculated lifetime crosses the characteristic hypernuclear weak-decay
timescale, \(\tau_{\mathrm{weak}}\simeq10^{-10}~\mathrm{s}\), near
\(m_H\simeq2020~\mathrm{MeV}\).  Below this mass, weak decay of the parent
double-\(\Lambda\) hypernucleus is faster than its strong conversion into
final states containing an \(H\).

In the representative dark-matter-motivated interval
\(1865\leq m_H\leq1885~\mathrm{MeV}\), the inclusive lifetime decreases from
\(6.80\times10^{-3}~\mathrm{s}\) at \(m_H=1865~\mathrm{MeV}\) to
\(6.59\times10^{-4}~\mathrm{s}\) at \(m_H=1885~\mathrm{MeV}\).  These values
exceed the characteristic weak-decay timescale by approximately seven to
eight orders of magnitude.  Furthermore, at \(m_H=1935~\mathrm{MeV}\), the
upper boundary of the mass region favored in our previous astrophysical
analysis, we obtain
\(\tau_{\mathrm{inc}}=1.93\times10^{-6}~\mathrm{s}\), which is still about
\(1.9\times10^{4}\) times longer than \(\tau_{\mathrm{weak}}\).
Consequently, although the breakup channels substantially strengthen the
strong conversion relative to the intact-\(\alpha\) calculation, the
observation of weakly decaying double-\(\Lambda\) hypernuclei does not
exclude a deeply bound \(H\) in these low-mass regions.

The absolute lifetime and the precise crossing mass remain sensitive to the
high-momentum behavior of the hypernuclear wave function.  Short-range
correlations can generate non-Gaussian momentum-space tails that decrease
more slowly than a single Gaussian.  Such components would generally
enhance the relevant overlap, increase the strong width, and shorten the
lifetime, potentially at the order-of-magnitude level.  Because the
high-momentum structure enters both the intact-\(\alpha\) normalization and
the breakup amplitudes, this effect cannot be included consistently through
a simple rescaling of the breakup contribution alone.  A correlated
calculation of the initial hypernuclear state and the final cluster
continuum is required to determine its quantitative impact.  A correction
of the expected magnitude could shift the model-dependent crossing near
\(m_H\simeq2020~\mathrm{MeV}\), but it would not overturn the low-mass
conclusion, where the calculated separation from the weak-decay timescale
spans several orders of magnitude.

Further refinements should include realistic short-range correlations,
antisymmetrized cluster-continuum states, 
%cluster spectroscopic amplitudes,
Coulomb effects, and strong final-state interactions.  The present results
nevertheless establish the central qualitative point: core breakup must be
considered when inferring an inclusive \(H\)-production rate from
double-\(\Lambda\) hypernuclei, while the resulting strong-conversion
lifetimes remain compatible with a deeply bound \(uuddss\) state in the
mass region discussed for sexaquark dark matter.  This conclusion concerns
the competition between strong \(H\) production and weak decay of the
parent hypernucleus; the lifetime of the \(H\) dibaryon itself is a separate
question not addressed here \cite{Gal2024}.
%Establishing a quantitative lifetime requires a fully antisymmetrized few-body treatment with realistic cluster wave functions and final-state interactions.

\section*{Acknowledgments}
M.S. acknowledges the hospitality of the Few-body Systems in Physics
Laboratory at the RIKEN Nishina Center and thanks Emiko Hiyama for the
invitation. M.S. also acknowledges support from the National Science
Centre, Poland (NCN), through SONATINA 7 Grant
No.~2023/48/C/ST2/00297. M.O. acknowledges partial support from the JSPS
Grants-in-Aid for Transformative Research Areas (Quantum Matter Science
in the Universe Opened Up by Precise Numerical Calculations),
JP25A203 and JP25H01267, and from the JSPS Grants-in-Aid for Scientific
Research JP23K03427.

\bibliographystyle{unsrtnat}
\begingroup
\small
\bibliography{references}
\endgroup

\end{document}